\documentclass[10pt,a4paper]{article}
\usepackage{graphicx}
\DeclareGraphicsExtensions{.jpg, .pdf, .png}

\date{}

\begin{document}

\date{}

\title{Basic cosmology\\
{\small Dedicated to Halton C. Arp}}
\author{Ll. Bel\thanks{e-mail:  wtpbedil@lg.ehu.es}}

\maketitle

\begin{abstract}

Basic cosmology describes the universe as a Robertson-Walker model filled with black-body radiation and no barionic matter, and as observational data it uses only the value of the speed of light, the Hubble and deceleration parameters and the black-body temperature  at the present epoch. It predicts the value of the next new parameter in the Hubble law.
\end{abstract}

{\it The Robertson-Walker model.}

\vspace{0.5cm}
Its line-element is:

\begin{equation}
\label{R-W}
ds^2=-dt^2+\frac{1}{c^2}\left(\frac{dr^2}{1-kr^2}+r^2(d\theta^2+\sin^2\theta d\phi^2)\right)
\end{equation}
where $c\equiv c(t)$ with $c_0=299792458.0$ m/s is the speed of light at the present epoch, ($t=0$)\,\footnote{$f(t)$ being a function  of $t$, $f_0$  means $f(0)$}.

$T\equiv T(t)$, with $T_0=2.74 K$ being the temperature of the  black-body radiation, the mass density $\rho\equiv \rho(t)$ and pressure $P\equiv P(t)$ are:

\begin{equation}
\label{rho-P}
\rho=\frac{a}{c^2}T^4, \quad  P=\frac{1}{3}\rho c^2
\end{equation}
where $a=7.565767\,10^{-16}$ J/m${}^3$/K${}^4$ is the radiation constant. No other mass contributes to the dynamics of the model. I assume that either it can be neglected because of its particular fractal distribution or because its content being highly unreliable now it is advisable to postpone its consideration to a non basic model.

Under the preceding assumptions Einstein's equations:

\begin{equation}
\label{Einstein}
R_{\alpha\beta}-\frac{1}{2}Rg_{\alpha\beta}-\Lambda g_{\alpha\beta}=-\frac{8\pi G}{c^2}(\rho c^2 u_\alpha u_\beta+P (g_{\alpha\beta}+u_\alpha u_\beta))
\end{equation}
where $\Lambda$ is the cosmological constant with dimensions T${}^{-2}$, reduce to the following two equations:

\begin{equation}
\label{Equ1}
Eq_1\equiv 3\dot{c}^2+3kc^4-\Lambda c^2=8\pi G\rho c^2
\end{equation}
and:

\begin{equation}
\label{Equ2}
Eq_2\equiv 2c\ddot{c}-5\dot{c}^2-kc^4+\Lambda c^2=8\pi GP
\end{equation}
Using (\ref{rho-P}) the following linear combination $1/2(Eq_1-3Eq_2)=0$ becomes:

\begin{equation}
\label{Equ3}
Eq_3\equiv c\ddot{c}-3\dot{c}^2-kc^4+\frac{2}{3}\Lambda c^2=0
\end{equation}
an equation that will be a useful substitute of $Eq_2$ once $\Lambda$ and $k$ will be known.

Solving the system of algebraic equations (\ref{Equ1}) and (\ref{Equ2}) with unknowns $\Lambda$ and $k$ we get:

\begin{equation}
\label{Lambda}
\Lambda=\frac{4\pi G}{c^2}(\rho c^2+3P)-\frac{3}{c^2}(c\ddot c-2\dot c^2)
\end{equation}
and:

\begin{equation}
\label{k}
k=\frac{4\pi G}{c^4}(\rho c^2+P)-\frac{3}{c^4}(c\ddot c-\dot c^2)
\end{equation}


{\it Hubble's law.}

\vspace{0.5cm}
Light emitted at time $t_e$ at some point with coordinate $r_e$ propagating in the radial direction and being received at time $t_r$ at a point with radial coordinate $r_r$ travels a proper distance $L$:

\begin{equation}
\label{L}
L=\int_{r_e}^{r_r}\frac{dr}{\sqrt{1-kr^2}},  \ \ \hbox{or}\ \ L=-\int_{t_r}^{t_e} c(t)\,dt,
\end{equation}
If $\delta t_e$ is the period of the light emitted at $r_e$ and $\delta t_r$ is the period of the light received at $r_r$ then from the preceding formula it follows that:

\begin{equation}
\label{variation}
0=c_r\delta t_r-c_e\delta t_e
\end{equation}
from where it follows that defining the red shift $z$ by:

\begin{equation}
\label{red-t}
z=\frac{\delta t_r}{\delta t_e}-1
\end{equation}
we get:

\begin{equation}
\label{z}
z=\frac{c_e}{c_r}-1
\end{equation}

The integral defining $L$ can be approximated as\,\footnote{Dots overhead mean derivatives with respect to $t$.}:

\begin{equation}
\label{approxL}
L=-c_r(t_e-t_r)-\frac12\dot c_r(t_e-t_r)^2 -\frac16\ddot c_r(t_e-t_r)^3
\end{equation}
and similarly $c_e$ can be approximated by:

\begin{equation}
\label{approxc}
c_e=c_r+\dot c_r(t_e-t_r)+\frac12\ddot c_r(t_e-t_r)^2 +\frac16\dot{\ddot c}_r(t_e-t_r)^3,
\end{equation}
Inverting (\ref{approxL}) we get:

\begin{equation}
\label{Invert}
t_e-t_r=-\frac{1}{c}L-\frac12\frac{\dot c}{c^3}L^2 +\frac16\frac{c\ddot c-3{\dot c}^2}{c^5}L^3
\end{equation}
and substitution of (\ref{Invert}) in (\ref{approxc}) and the result in (\ref{z}) we get:

\begin{equation}
\label{Invert2}
z=-\frac{\dot c}{c^2}L+\frac12\frac{c\ddot c-\dot c^2}{c^4}L^2+\frac16\left(\frac{4c\dot c\ddot c-3\dot c^3-c^2\dot{\ddot c}}{c^6}\right)L^3
\end{equation}

Defining the Hubble function $H$ and the deceleration parameter $q$ as usual, and the jerk parameter $j$\,\footnote{$j=a^2\dot{\ddot a}/\dot a^3$ if $a=c_0/c$ is the scale factor.}

\begin{equation}
\label{Hq}
H=-\frac{\dot c}{c}, \ \  q=\frac{c\ddot c}{\dot c^2}-2, \ \ j=6-6\frac{c\ddot c}{\dot c^2}+\frac{c^2\dot{\ddot c}}{\dot c^3}
\end{equation}
we extend with an extra term the well known Hubble formula:

\begin{equation}
\label{basic law}
z=\frac{H}{c}L +\frac12\frac{H^2}{c^2}(1+q)L^2+\frac16\frac{H^3}{c^3}(6+6q+j)L^3
\end{equation}

The formulas (\ref{Lambda})) and (\ref{k}) can be written:

\begin{equation}
\label{basic Lambda}
\Lambda=\bar\Lambda+4\pi G\left(\rho+\frac{3P}{c^2}\right), \ \hbox{where}\  \bar\Lambda=-3H^2q
\end{equation}
and:

\begin{equation}
\label{basic k}
k=\bar k+\frac{4\pi G}{c^2}\left(\rho+\frac{P}{c^2}\right)  \ \hbox{where}\ \bar k=-\frac{H^2(1+q)}{c^2}
\end{equation}


{\it Observational data}
\vspace{0.5cm}

The Hubble constant and the deceleration parameter have been measured to be $H_0=72$ km/s/Mpc and  $q_0=-0.55$.
This is all that is needed with $c_0$ and $T_0$ to derive the values of the cosmological constant $\Lambda$ and the curvature constant $k$  using Eqs. (\ref{basic Lambda}) and  (\ref{basic k}).

For $t=0$  the r-h-s of these two equations are known:

\begin{equation}
\label{Lambda-k0}
\bar\Lambda_0= 8.983543533\,10^{-36}\,s^{-2}, \ \ \bar k_0=-2.726056423\,10^{-53}\,m^{-2}
\end{equation}

and:

\begin{equation}
\label{rho0-P0}
\rho_0=4.473697482\,10^{-31}\,kg\, m^{-3}, \ \ P_0=1.340252927\,10^{-14}\,kg\, m^{-1}\, s^{-2}
\end{equation}

and therefore the constants $\Lambda$ and $k$ are:

\begin{equation}
\label{Lambda-k}
\Lambda=8.984293774\,10^{-36}\,s^{-2}, \ \ k=-2.725499919\,10^{-53}\,m^{-2}
\end{equation}

With $\Lambda$ and $k$ known, from (\ref{Equ3}) we get:

\begin{equation}
\label{D2c}
\ddot c=\frac13\left(\frac{3c^4k+9\dot c^2 -2c^2\Lambda}{c}\right)
\end{equation}
Differentiating now with respect to $t$ we have:

\begin{equation}
\label{D3c}
\dot{\ddot c}=\dot c\left(9c^2k+\frac{15\dot c^2}{c^2}-\frac{14}{3}\Lambda\right)
\end{equation}
and using now the definitions (\ref{Hq}) we find the following convenient expression for the new function:

\begin{equation}
\label{D3c0}
j=\frac13\left(\frac{9H^2+9c^2k-2\Lambda}{H^2}\right)
\end{equation}
Therefore the predicted observational value of $j(0)$ is\,\footnote{this value is quite compatible with one of the determinations of $j_0=0.631\pm 0.290$ mentioned in {\bf Table} 3 of reference \cite{Akarsu}}:

\begin{equation}
\label{q0}
j(0)=0.55
\end{equation}


{\it Maximally symmetric models\,\footnote{See \cite{Weinberg} for instance} }
\vspace{0.5cm}

Let us assume that at some value of $t$ we know the values of $c$, $\dot c$ and $\ddot c$ or equivalently, from (\ref{Hq}), the values of $c$, $H$ and $q$ corresponding to some general function $c(t)$. These data are sufficient to calculate the Riemann, and Einstein's tensors of the the line-element (\ref{R-W}) at this time $t$. I call Osculating model, \cite{Bel0}, at time $t$ the Robertson-Walker model with line-element:

\begin{equation}
\label{R-W0}
ds^2=-dt^2+\frac{1}{\bar c^2}\left(\frac{dr^2}{1-\bar kr^2}+r^2(d\theta^2+\sin^2\theta d\phi^2)\right)
\end{equation}
solution of Einstein's equations:

\begin{equation}
\label{Einstein0}
R_{\alpha\beta}-\frac{1}{2}Rg_{\alpha\beta}-\bar\Lambda g_{\alpha\beta}=0
\end{equation}
where $\bar\Lambda$ and $\bar k$ are defined in (\ref{basic Lambda}) and (\ref{basic k}). This is a vacuum solution but it is more than that: it is one of the space-times with maximum symmetry. This meaning that the Riemann tensor is:

\begin{equation}
\label{Riemann0}
R_{\alpha\beta\mu\nu}=-\frac13\bar\Lambda(g_{\alpha\mu}g_{\beta\nu}-g_{\alpha\nu}g_{\beta\mu})
\end{equation}
and therefore it is invariant under a 10 dimensional  group of isometries.

The concept of osculating model allow us to look at the history of the Universe as a continuous unfolding of maximally symmetric space-times while we live at an epoch when our osculating universe is one of de Sitter's space-time models, type $dS_-$, with:

\begin{equation}
\label{barc}
\bar c=\frac{\bar\lambda_0}{\bar p_0}\hbox{csch}\left(\bar\lambda_0 t+\hbox{csch}^{-1}\left(\frac{\bar p_0c_0}{\bar\lambda_0}\right)\right), \ \bar\lambda_0=\sqrt{\frac{\bar\Lambda_0}{3}}, \ \bar p_0=\sqrt{-\bar k_0}
\end{equation}
and the values of $\Lambda$ and $k$ are  so close to the values of $\bar\Lambda_0$ and $\bar k_0$ that the functions $c$ and $\bar c$ are almost undistinguishable in the interval $t=-1..0$.
The Figures 1-4 are  in succession the graphs of $c$, $\rho$, $\bar\Lambda$ and $\bar k$. The units are such that: $c_0=1$, $H_0=1$
$\rho_0=1.4\,10^{-4}$, $\Lambda_0=1.65$, $k_0=-0.45$ and $8\pi G=1$.
\vspace{5mm}


{\it Hamiltonian formalism}
\vspace{5mm}

The differential equation (\ref{Equ3}) describes the ensemble ${\cal E}(\Lambda,k)$ of Robertson-Walker models (\ref{R-W}) when its source is such that:

\begin{equation}
\label{P-rho}
P=\frac13\rho c^2
\end{equation}
It can be derived from the Lagrangian:

\begin{equation}
\label{Lagrangian}
{\cal L}=\frac12\frac{\dot c^2}{c^6}-\frac12\frac{k}{c^2}+\frac16\frac{\Lambda}{c^4}
\end{equation}
whose associated Hamiltonian is the constant of motion:

\begin{equation}
\label{Hamiltonian}
{\cal H}=\frac12\frac{\dot c^2}{c^6}+\frac12\frac{k}{c^2}-\frac16\frac{\Lambda}{c^4}, \ \ \ \ \frac{d{\cal H}}{dt}=0
\end{equation}
To each value of ${\cal H}$ corresponds a sub-ensemble of models ${\cal S}(\Lambda,k; {\cal H})$, and if in particular ${\cal H}=0$ then ${\cal S}(\Lambda,k; 0)$ is the ensemble of maximally symmetric space-time models. Therefore somehow the value of ${\cal H}$ of a model of the ensemble ${\cal E}(\Lambda,k)$ measures its deviation with respect to  the corresponding maximally symmetric model.

Using the data and the units mentioned at the end of the preceding section the value of ${\cal H}$ for the basic universe described in this paper is $0.005$.

From (\ref{Equ1}) and (\ref{Hamiltonian}) it follows the eventually useful formula:

\begin{equation}
\label{H-rho}
\rho=\frac34 \frac{c^4{\cal H}}{\pi G}
\end{equation}

\vspace{5mm}

{\it Open question}

\vspace{5mm}
Basic cosmology assumes that light from stars and galaxies propagates freely across the black-body radiation. But since it is known that while there is no direct photon-photon interaction there are indirect interactions through intermediate virtual particles \cite{Enterria}, it could be\,\footnote{More on that in \cite{Bel}}
that the black body fluid has an index of refraction; and in this case light would not propagate along null geodesics of the Robertson-Walker model and cosmology would radically change from what now we believe it to be.

\begin{figure}[htbp]
\begin{center}
\begin{minipage}[b]{15cm}
\includegraphics[width=5cm]{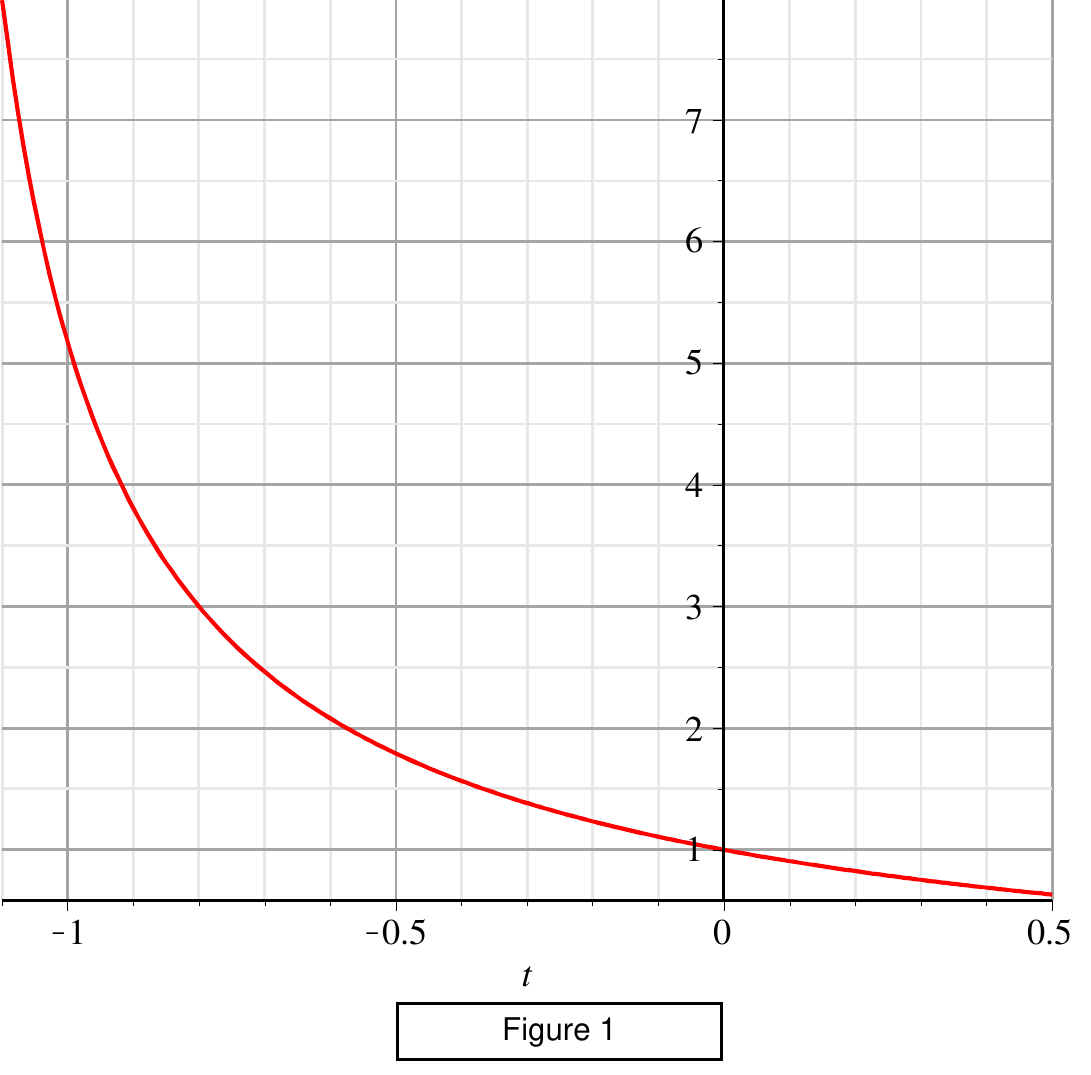}
\includegraphics[width=5cm]{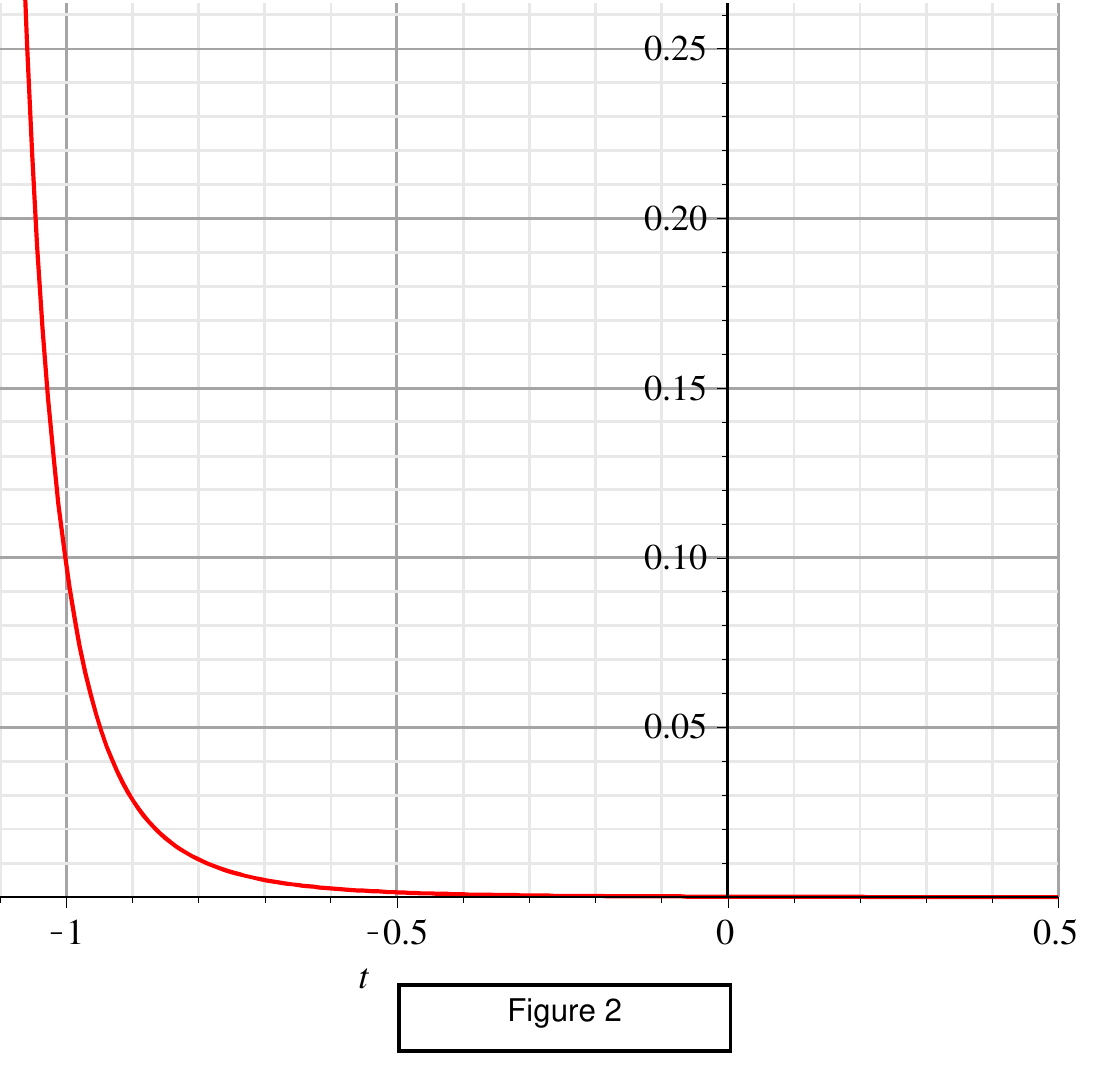}
\end{minipage}
\end{center}
\end{figure}

{\bf Figure} 1 ($c$), {\bf Figure} 2 ($\rho$),

\begin{figure}[htbp]
\begin{center}
\begin{minipage}[b]{15cm}
\includegraphics[width=5cm]{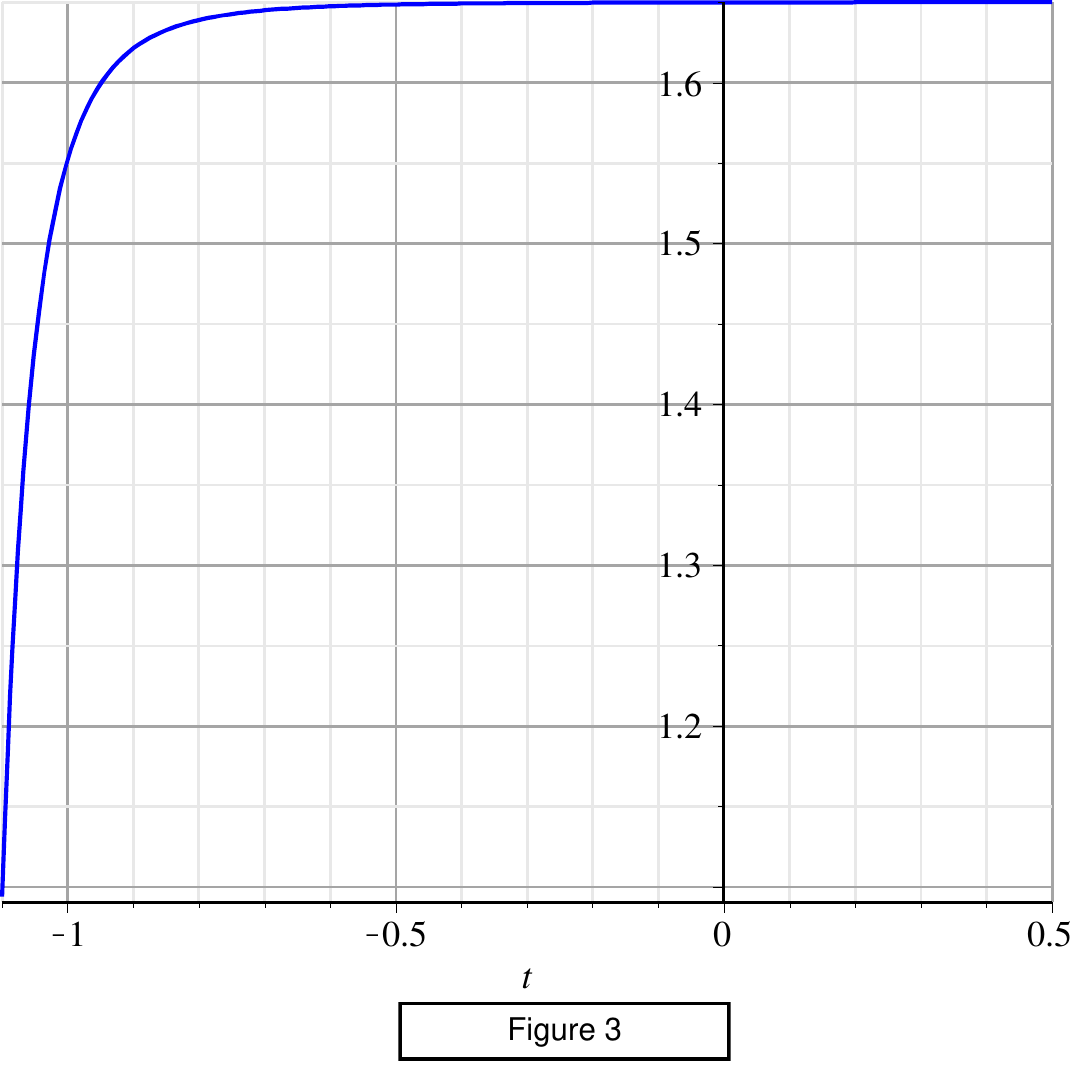}
\includegraphics[width=5cm]{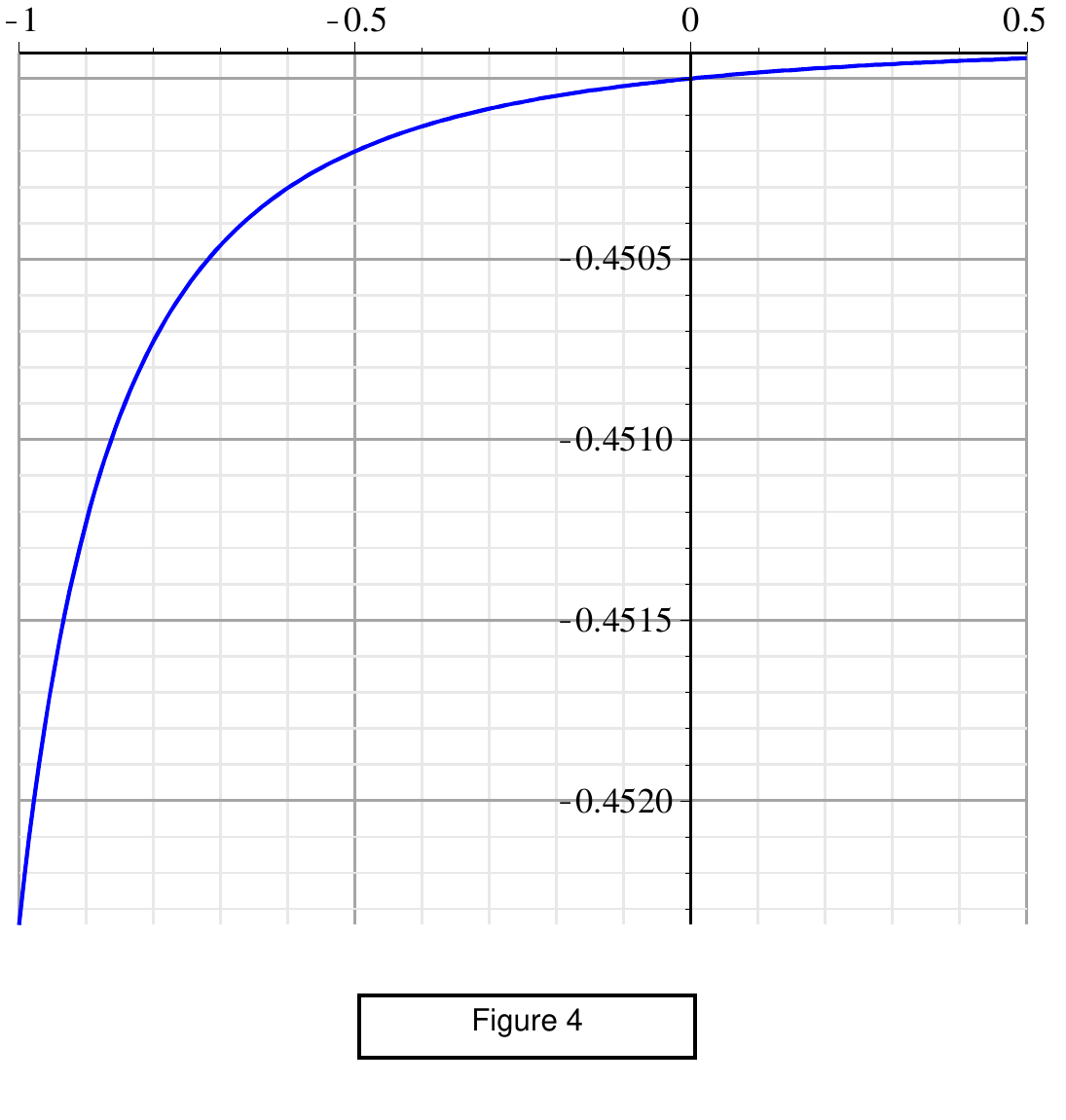}
\end{minipage}
\end{center}
\end{figure}

{\bf Figure} 3 ($\bar\Lambda$), {\bf Figure} 4 ($\bar k$).

 \end{document}